\documentclass[10pt,a4paper]{article}

\usepackage{epsfig}
\usepackage{amsmath}
\usepackage{graphicx}
\usepackage{float}
\usepackage{subfig}
\usepackage{vmargin}
\usepackage{mathrsfs}
\usepackage{mathbbol}
\usepackage{hyperref}

\newcommand{\be}{\begin{equation}}
\newcommand{\ee}{\end{equation}}
\newcommand{\bes}{\begin{equation*}}
\newcommand{\ees}{\end{equation*}}
\newcommand{\ba}{\begin{equation} \begin{aligned}}
\newcommand{\ea}{\end{aligned} \end{equation}}

\newcommand{\SIR}{{\it SIR}}

\title{Modelling Epidemics on Networks}
\author{Thomas House, Warwick Mathematics Institute}
\date{}
\newfont{\mtfont}{percsymbols}
\newcommand{\mtIi}{\mbox{\mtfont B}}
\newcommand{\mtIii}{\mbox{\mtfont C}}
\newcommand{\mtIiii}{\mbox{\mtfont D}}
\newcommand{\mtIiv}{\mbox{\mtfont E}}
\newcommand{\mtIIi}{\mbox{\mtfont F}}
\newcommand{\mtIIii}{\mbox{\mtfont G}}
\newcommand{\mtIIiii}{\mbox{\mtfont H}}
\newcommand{\mtIIiv}{\mbox{\mtfont I}}
\newcommand{\mtII}{\mbox{\mtfont J}}

\begin{document}

\maketitle

\begin{abstract}
Infectious disease remains, despite centuries of work to control and mitigate
its effects, a major problem facing humanity. This paper reviews the
mathematical modelling of infectious disease epidemics on networks, starting
from the simplest Erd\"{o}s-R\'{e}nyi random graphs, and building up structure
in the form of correlations, heterogeneity and preference, paying particular
attention to the links between random graph theory, percolation and dynamical
systems representing transmission. Finally, the problems posed by networks with
a large number of short closed looks are discussed.
\end{abstract}

\section{Introduction}

\label{sec:intro}

In science (and particularly theoretical science) it is necessary to
approximate to make progress. Such approximations can be non-rigorously
categorised into three types.
\begin{enumerate}
\item[Type I] Approximations that cut out the `unnecessary' complexity in a system
to yield an appropriate mathematical representation. For example, it is not
necessary to worry about the weak nuclear force when you are modelling a bungee
jump; gravity and the elasticity of the rope are all that is needed.
\item[Type II] Approximations that are controlled up to a power of some small
quantity, i.e.\ ``Corrections to this result will be $O(\varepsilon{})$.''
\item[Type III] Approximations of mathematical convenience, made so that a system
can be analysed but with little other motivation.
\end{enumerate}
As a former physicist now working in biology, I believe that the reason physics
has been so successful in developing theory is that physical experimental
systems can often be modelled by making Type I and Type II approximations. In
biology, medicine, and sociology, it is much more frequent that Type III
approximations are made.

Despite this difficulty in making well motivated simplifying assumptions,
modern biology is one of the most exciting areas of science, with enormous
quantities of interesting experimental data that pose major unanswered
scientific questions. In some cases, laboratory techniques are able to make
relatively precise measurements -- although these are nowhere near the accuracy
of, say, atomic physics -- but in other cases even repeatable experiments are
not possible.  This is similar to what happens in cosmology, where we only see
one realisation of the universe. Cosmological models are, of course, highly
informative and useful, but they are not as accurate as the incredible
agreement between theory and data on the spectrum of atomic hydrogen.

Epidemiology is the study of patterns of disease in populations. It started out
as the study of infectious agents (the topic of this review) but has grown to
encompass the diseases of lifestyle and affluence like cancer and heart
disease.  Sometimes controlled experiments infect human volunteers with mild
illnesses, or cause more severe disease in non-human animals, but since
infectious diseases can never be ethically released into non-laboratory
populations, infectious disease epidemiology as defined is an observational
science, like cosmology.  While life-threatening infections are now largely
under control in rich countries, they kill millions, including a large
proportion of children under 5, in the developing world~\cite[Fig.\
5]{WHO:2008}. One of the global problems we face in the 21st century is the
unequal distribution of calorie intake.  In the US and EU, obesity is driving
epidemics of heart disease and diabetes; but malnutrition is a large part of
the reason that infectious diseases are so deadly for poor
children~\cite{WHO:2008}.

While we wait for a political solution to the fundamental problem of global
inequality, scientists can do several things to help limit the human cost of
infectious diseases. Laboratory biologists and clinicians can develop and test
new treatments, vaccines and behavioural interventions. But given the possibility
of deploying such interventions, there will always be the question of how to
optimise their use. Epidemiologists have therefore got two scientific tasks:
first, to identify the key routes of disease transmission; and secondly, to
design optimal interventions making use of that knowledge.

The aim of this review is to introduce some of the mathematics used in modern
infectious disease epidemiology, in particular the sub-discipline of modelling
epidemics on networks, where tools from statistical physics are increasingly
used. In contrast to much of physics, there is very little consensus
amongst researchers about many important issues in epidemic modelling. This
means that a review must either be highly technical, so that a reader can make
up their own mind, or somewhat subjective. I have adopted the latter approach,
trying to signpost clearly where a statement is a personal opinion. Also, key
biological insights obtained from these mathematical techniques are
highlighted at appropriate points.

\section{The \SIR{} model}

\subsection{Infection dynamics}

Suppose we have a `closed population' -- i.e.\ a large number of individuals,
with no births or deaths during the period of time modelled. To motivate the
model most commonly considered in epidemic models, three approximations are made.
\begin{enumerate}
\item The \emph{compartmental} approximation: Each individual is either susceptible
to infection, infectious with the disease, or recovered and immune. Write $S(t)$
for the proportion of the population that is susceptible, $I(t)$ for the
proportion that is infections, and $R(t)$ for the proportion that is recovered.
\item The \emph{mass-action} approximation: Infection happens between each
susceptible and infectious individual in the population at a constant rate $\beta$.
\item The \emph{Markovian} approximation: Recovery from infection is a Markovian
(memory-less) process, and happens at a constant rate $\gamma$.
\end{enumerate}
These assumptions, lead to the \SIR{} (susceptible-infectious-recovered)
equations below in the limit as population size becomes extremely large.
\ba
\frac{dS}{dt} & = - \beta S I \text{ ,}\\
\frac{dI}{dt} & = \beta S I -\gamma I \text{ ,}\\
\frac{dR}{dt} & = \gamma I \text{ .} \label{sir}
\ea
It is worth thinking briefly about the approximations introduced. I would argue
that, at least for some diseases, the compartmental assumption is Type I -- it
helps us to get our thinking about the problem straight, even if there is a
more detailed microscopic story about microbes, white blood cells and
antibodies. Putting people into discrete compartments is what empirical
epidemiologists often do, counting present and former cases rather than trying
to determine where all the viruses and bacteria are. In contrast, the Markovian
approximation is definitely Type III. There is no good reason to think that
recovery from illness has no memory and is as likely one hour after infection
as it is a week after. But Markovian dynamics remain popular for two reasons:
many results are not sensitive to this assumption; and it does massively
simplify epidemic modelling. Finally, the mass-action assumption actually works
well for small, well-connected populations like boarding schools~\cite[Fig.\
2.4]{Keeling:2007}; but actually the equations~\eqref{sir} assume that the
population is extremely large. I would therefore categorise mass action,
alongside the Markovian approximation, as something that is assumed for
mathematical convenience.

\subsubsection*{Biological insight}

\begin{figure}
\begin{center}
\begin{minipage}{60mm}
\resizebox*{6cm}{!}{\includegraphics{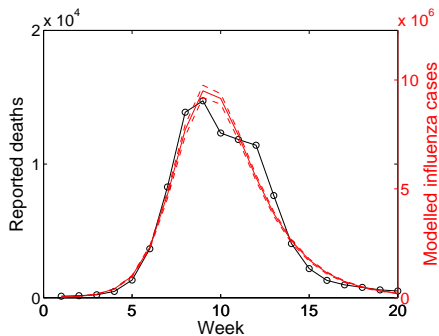}}%
\caption{Comparison of the \SIR{} model (solid red line, with 95\% CI as dashed
red lines) with death data (black line with circles) for the main wave of the
1918-19 influenza pandemic in England and Wales.}%
\label{fig:1918}
\end{minipage}
\end{center}
\end{figure}

Despite its simplicity, the \SIR{} model fits many epidemics well.
Figure~\ref{fig:1918} shows what happens when this model is fitted to the main
wave of the 1918-19 pandemic in England and Wales. Clearly, there are features of
the data that are not captured by the model, but it still acts as a good
`starting point' for understanding epidemics of infectious disease.

\subsection{Early behaviour of an epidemic}

Network theory enters epidemiology as an attempt to relax the assumption of mass
action; before turning to this, let us analyse equations~\eqref{sir} using the
theory of dynamical systems. The first thing to note about them is that they
are \emph{conservative}, i.e.\ the quantity $S+I+R$ is invariant over time.
This follows from the assumption that there are no births and deaths in the
population, and means that we only need to specify two initial conditions to
integrate the system: $S(0)$ and $I(0)$ such that $S(0) + I(0) \leq 1$. If we
start out with $I(0) \ll 1$, then the equations~\eqref{sir} can be linearised
to give
\be
\frac{dI}{dt} = \beta S(0) I - \gamma I + O\left(I^2\right) 
\quad \Rightarrow \quad
I(t) \approx I(0) e^{(\beta S(0) - \gamma) t} \text{ .}
\label{linsir}
\ee
Therefore, if a small amount of infection is introduced into a population,
we will see initial exponential growth if $\beta S(0) > \gamma$, and a decline
in the number of infectious individuals otherwise.

\subsection{The basic reproductive ratio}

A quantity that is often defined in epidemic models is the \textit{Basic
reproductive ratio}, $R_0$ (distinct from the initial proportion of the
population in the recovered group, $R(0)$, in an unfortunate but fixed
notational convention). This is defined verbally as the expected number of
additional infectious individuals produced by a typical infectious individual
early in the epidemic. By simple logical argument, this quantity must exceed
unity for an epidemic to grow, since otherwise each infected fails to produce, on
average, more than one new infected before they recover. For the \SIR{} model
above, we can simply write down the basic reproductive ratio:
\be
R_0 = \frac{\beta S(0)}{\gamma} \text{ .}
\ee
Clearly, the verbal argument that this quantity should exceed unity for an
epidemic to take off agrees with the dynamical argument made above about the
linearised system~\eqref{linsir}. $R_0$ is widely regarded as one of the most
important contributions of mathematical analysis to infectious disease
epidemiology, and can be defined for many different epidemic
models~\cite{Diekmann:2000}. On a general network, the appropriate definition
of this quantity becomes more difficult; instead it is easier to focus on early
behaviour, as analysed above, and epidemic final size.

\subsection{Final size and vaccination}

Now let us manipulate the \SIR{} model~\eqref{sir}, dividing the first equation
by the third to give
\be
\frac{dS}{dR} = - \frac{\beta}{\gamma} S
\quad \Rightarrow \quad
S(t) = S(0) e^{(R(0) - R(t) )\beta / \gamma} \text{ .}
\label{sreq}
\ee
This relationship allows us to derive results about the final outcome of an
epidemic. Before doing this, let us consider the impact of vaccination on the
infection. Conceptually, there are two kinds of vaccination that represent
extreme limits within which biological reality falls. The first of these is
\textit{leaky vaccination}, which reduces the susceptibility of every
individual in the population. This is modelled by scaling the transmission rate
$\beta \rightarrow \varepsilon \beta$ for $\varepsilon$ between 0 and 1. The
second kind of vaccination is called \textit{all-or-nothing vaccination}, in
which a proportion $p_V$ of the population is vaccinated and completely immune
to disease. This is modelled by taking initial conditions $S(0)=1 - p_V-I(0)$,
$I(0) \ll 1$, $R(0) = p_V$. In reality, vaccines can provide comprehensive
immunity in some individuals and partial immunity in others, while coverage
never reaches 100\%; the concepts of leaky and all-or-nothing vaccination are
therefore best seen as limiting cases.

Having included the effects of vaccination, we can then use the
result~\eqref{sreq} to calculate the value $R(\infty)-p_V$, which is the
proportion of a population experiencing disease during an epidemic, a quantity
often called the \textit{attack rate} by epidemiologists.  Note that medics 
use the word `rate' incorrectly to mean a proportion or percentage, rather than
something with units of time$^{-1}$. They are not going to stop doing this, so
it is just something quantitative scientists have to live with. The attack
rates calculated are shown in Figure~\ref{fig:sir-er}(a).
 
\subsubsection*{Biological insight}

\begin{figure}
\begin{center}
\begin{minipage}{120mm}
\subfloat[]{
\resizebox*{6cm}{!}{\includegraphics{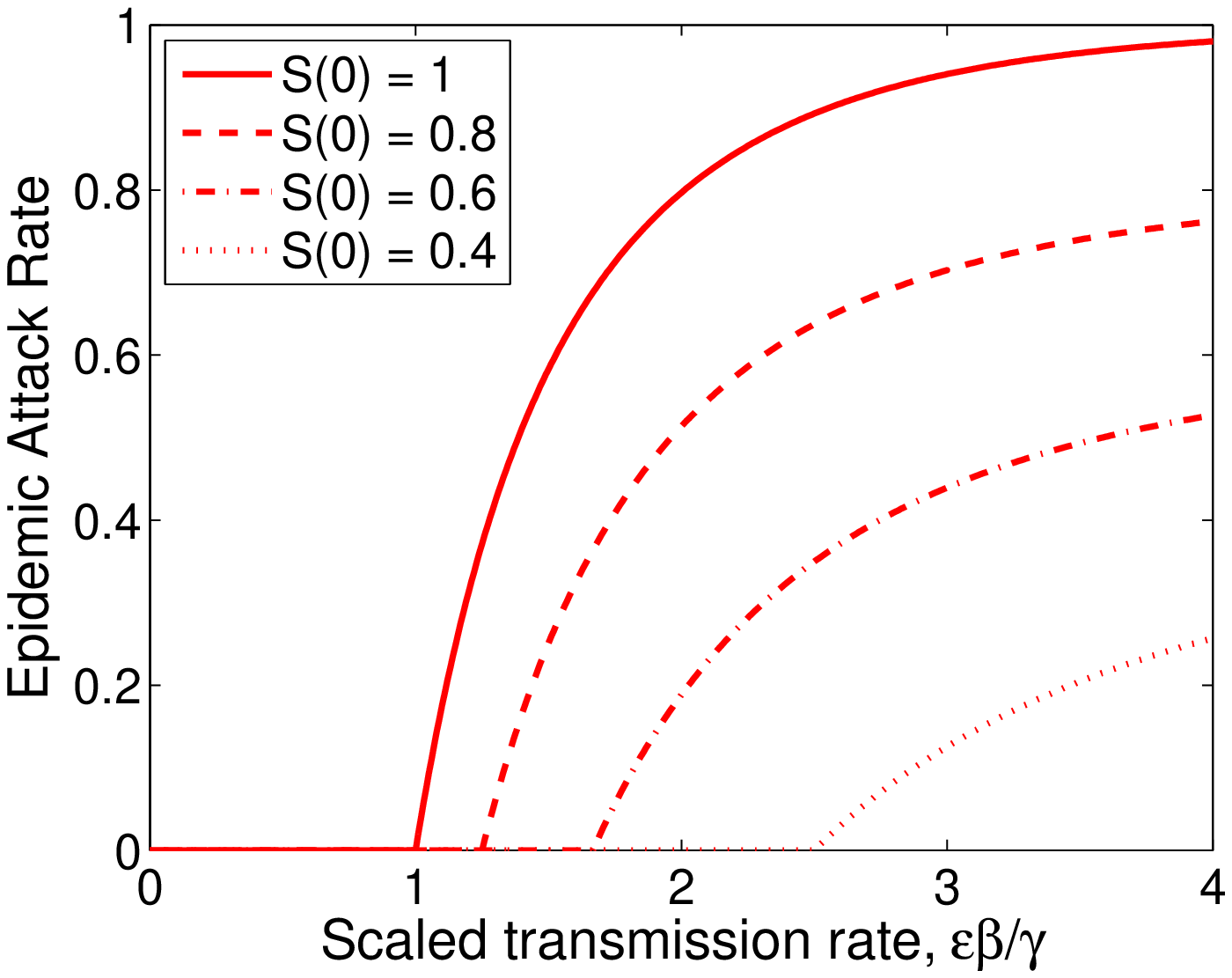}}}%
\subfloat[]{
\resizebox*{6cm}{!}{\includegraphics{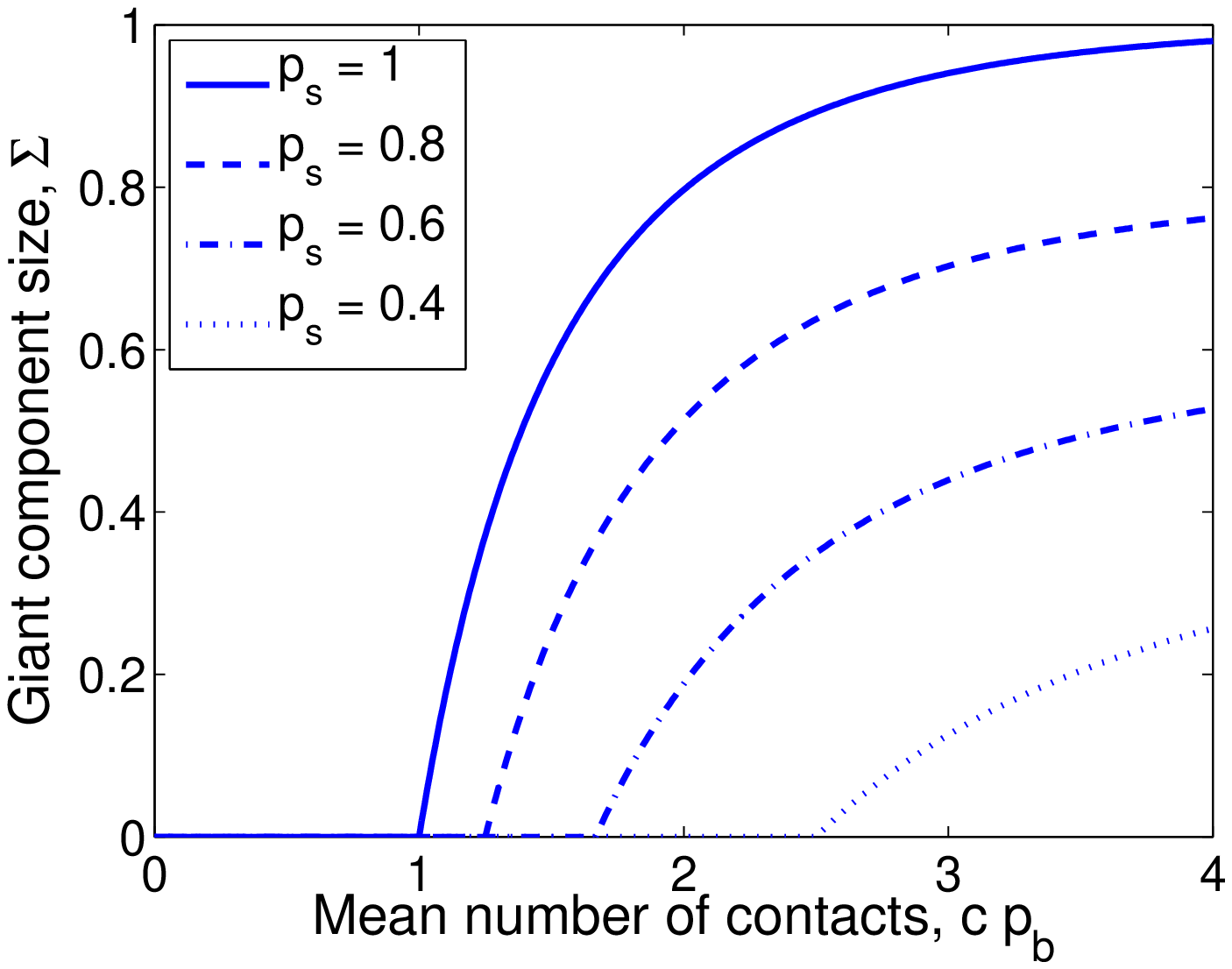}}}%
\caption{(a) shows the final size of an \SIR{} epidemic. Note that for this
model, $R_0 = \varepsilon \beta S(0) / \gamma$. (b) shows the giant component
size of an ER random graph -- spot the difference!}%
\label{fig:sir-er}
\end{minipage}
\end{center}
\end{figure}

There are several interesting features of Figure~\ref{fig:sir-er}(a), but three
are particularly worth highlighting:
\begin{enumerate}
\item A finite proportion of the population experiences disease if and only if
$R_0 > 1$. Below this threshold the final size is zero.
\item $R_0$ does not need to be much larger than 1 to generate an extremely
large epidemic. The gradient of the curves is steep for $R_0$ just over 1.
\item Regardless of how large the transmission rate is, the final size is always
strictly less than $S(0)$. Epidemics end because they run out of infectious
individuals, not because they run out of susceptibles.
\end{enumerate}
We now turn to the general theory of networks, before looking for parallels
between these and epidemics.

\section{Network theory}

\subsection{Fundamental concepts}

\label{sec:fcs}

A network is made up of two objects: \textit{nodes}, and \textit{links}. These
are abstract terms for the individuals we want to consider, and the
relationships between them respectively. In the current context, these could be
individuals who can be infected and contacts that can lead to the transmission
of disease. Suppose we have $N$ nodes labelled by indices $i,j = 1, \ldots, N$.
Then a common way to represent network structure is through the
\textit{adjacency matrix} $\mathbf{G} = (G_{ij})$, where
\be
G_{ij} = G_{ji} = \begin{cases}
1 & \text{ if } i \text{ and } j \text{ are linked,}\\
0 & \text{ otherwise.}
\end{cases}
\label{Gdef}
\ee
It is possible to define generalisations of this, where the matrix is not equal
to its transpose (leading to an \textit{asymmetric} network) or takes general
values (leading to a \textit{weighted} network). There is also the question of
whether a node can be connected to itself; but in the context of infectious
disease, it makes most sense to assume that $G_{ii}=0$ so that nodes do not
link to themselves.

There are many different properties of a network that can be defined, and a
recent textbook summarises these quite comprehensively~\cite{Newman:2010}.
Perhaps the most fundamental, however, is the notion of a node's
\textit{degree}. The degree of node $i$ is
\be
k_i = \sum_{j} G_{ij} \text{ .}
\ee
We will also write $N_k$ for the number of nodes of degree $k$, so that $d_k =
N_k /N$ is a discrete distribution known as the network's \textit{degree
distribution}.

Another particularly important concept for epidemic networks is that of a
\textit{component} -- a set of nodes for which any pair is linked to each other
through a finite-length path through the network. By labelling the nodes
correctly, it is possible to write the adjacency matrix in block diagonal form:
\be
\mathbf{G} = \left( \begin{array}{cccc}
\mathbf{G}_{(1)} & \mathbf{0} & \mathbf{0} & \cdots \\
\mathbf{0} & \mathbf{G}_{(2)} & \mathbf{0} & \cdots \\
\mathbf{0} & \mathbf{0} & \mathbf{G}_{(3)}  & \cdots \\
\vdots & \vdots & \vdots & \ddots
\end{array} \right) \text{ ,}
\ee
so that $\mathbf{G}_{(C)}$ is the adjacency matrix for component $C$. There is
clearly a qualitative difference between a network in which a significant
number of the nodes are in one component, and a network made up of many small
components.  We will now turn to how networks can move between one regime and the
other.

\subsection{Erd\"{o}s-R\'{e}nyi random graphs}

Stochastic processes that produce networks are called \textit{random graph
models}. The word `graph' is essentially synonymous with the word `network' in
this context, although some authors do make a distinction. These models are
useful for a variety of reasons. It may be that the family of networks produced
by a random graph model has interesting properties; or the random graph model
might be used as a null model -- i.e.\ something to test against real data --
in statistical work.

The Erd\"{o}s-R\'{e}nyi (ER) random graph model involves taking $N$
individuals, and putting a link between each of the $N(N-1)/2$ pairs of
individuals with independent probability $\pi$. While a highly mathematical
treatment of this model is possible~\cite{Durrett:2007}, we will argue
heuristically here. Of particular interest is the size of components the
network produced. The largest component in a network is called the
\textit{giant component}; the key qualitative difference of network types is
whether the giant component size as a proportion of the nodes $S$ tends to 0 as
$N\rightarrow\infty$, or whether it tends to some finite value between 0 and 1.

Let us suppose we are in the latter situation, and pick a random node in the
graph. The probability that this node is not in the giant component is $x$,
which is the same for all nodes since they are not differentiated and we have
picked randomly. Now consider all other nodes in the network -- if the initial
node is not in the giant component, then they must be either not connected to
the initial node, or connected to the initial node and not in the giant
component themselves. We can write this statement mathematically as
\be
x = \left((1-\pi) + \pi x\right)^{N-1} \text{ ,}
\label{biner}
\ee
which is a polynomial in $x$ with no simple analytic solution. As $N$
increases, even numerical solution of~\eqref{biner} becomes difficult, and it
is necessary to take the limit $N\rightarrow\infty$, holding constant the mean
number of contacts per node $c = (N-1)\pi$, so that
\be
x = e^{(x-1)c} \label{limer}
\ee
is the appropriate equation for the probability that a node is not in the giant
component of a very large ER graph.  Already, the similarity between this
expression and~\eqref{sreq} should be clear, but the analogy can be made still
stronger by consideration of a slightly more general model. Before doing this,
note that $c$ is the mean node degree, and the network's degree distribution
(as defined in \S{}\ref{sec:fcs} above) will be Poisson with parameter $c$ in
the limit $N\rightarrow\infty$.

One important qualitative feature of the ER random graph model is that it
undergoes a \textit{phase transition} at $c=1$. Below this critical value, the
equation~\eqref{limer} does not have a solution such that $0\leq x <1$ and there
is no sizeable giant component -- the network is made of lots of small
components. For $c>1$, one component dominates meaning that a disease spreading
around the population can reach a significant proportion of individuals.

\subsection{Percolation on graphs and epidemics}

Percolation is a standard tool in statistical physics~\cite{Stauffer:1994}; in
the context of networks, there are two ways that this method is used to modify
an existing network. In \textit{site percolation}, each original node is
present in the modified network with independent probability $p_s$ (i.e.\ we
remove a proportion $1-p_s$ of nodes at random). In \textit{bond percolation},
each original link is present in the modified network with independent
probability $p_b$ (i.e.\ we remove a proportion $1-p_b$ of links at random).

Suppose we have applied both node and link removals to an ER random graph, and
make the same argument about picking a node at random. If the node is not in
the giant component, then either it is not present following site percolation,
or else all other nodes in the network that survive site percolation must
either: (i) not be linked in the original ER graph, or be linked in the
original ER graph and have the link deleted during bond percolation; or (ii) be
linked in the original ER graph, have the link remain during bond percolation,
and not be in the giant component. After some mathematical manipulations along
the lines of those used to derive~\eqref{limer}, taking the $N\rightarrow\infty$ 
limit for constant $c$ gives
\be
x = (1-p_s) + p_s e^{(x+1-2p_s)c p_b} \text{ .}
\label{binperc}
\ee
Then the giant component size is given by $\Sigma=1-x$, once we have solved for
$x$.  Figure~\ref{fig:sir-er}(b) shows the results of doing this; having gone
through all the maths above, it is not surprising both plots in
Figure~\ref{fig:sir-er} are identical since the equations used to generate them
are mathematically isomorphic. In fact we can write this equivalence out
explicitly as shown in Table~\ref{equivtab}.  But if you had seen the \SIR{}
equations~\eqref{sir}, and heard a description of both the ER random graph
model and percolation, would this equivalence be obvious?

\begin{table}
\begin{center}
{\begin{tabular}{p{5cm}|p{5cm}}
\textbf{Epidemic} & \textbf{Network} \\
\hline 
Basic reproductive ratio, $R_0$ & Mean node degree, $c$ \\
Leaky vaccine scaling, $\varepsilon$ & Remaining fraction of links after
bond percolation, $p_b$ \\
All-or-nothing vaccination level, $p_V$ & Fraction of nodes removed by
site percolation, $1-p_s$ \\ 
Attack rate, $R_\infty - p_V$ & Giant component size, $\Sigma$ \\
\end{tabular}}
\end{center}
\caption{Parallels between \SIR{} epidemics and ER random graphs}
\label{equivtab}
\end{table}

I do not believe it would. In fact, the relationship between differential
equations that describe dynamical processes over time and static models based
on probabilities is quite subtle. For example, some infections do not lead to
long-lasting immunity and after recovery individuals can become susceptible
again. For such infections, there is no simple process of deletions or nodes
and links that gives the potential of the epidemic to take off or the final
impact of the epidemic. Similarly, even for the simpler \SIR{} case where
immunity to further infection is long-lasting, site- and bond percolation do
not work as calculational tools where short, closed loops are present in the
network in appreciable numbers.

But where percolation works, it is very useful, since there are few other
analytic approaches to epidemics on networks. Generally applicable Monte-Carlo
methods, where a computer picks random numbers to simulate the epidemic
process, can be highly computationally intensive. 

\section{Correlation, heterogeneity and preference in epidemic networks}

\begin{figure}
\begin{center}
\begin{minipage}{135mm}
\subfloat[]{
\framebox{\resizebox*{4cm}{!}{\includegraphics{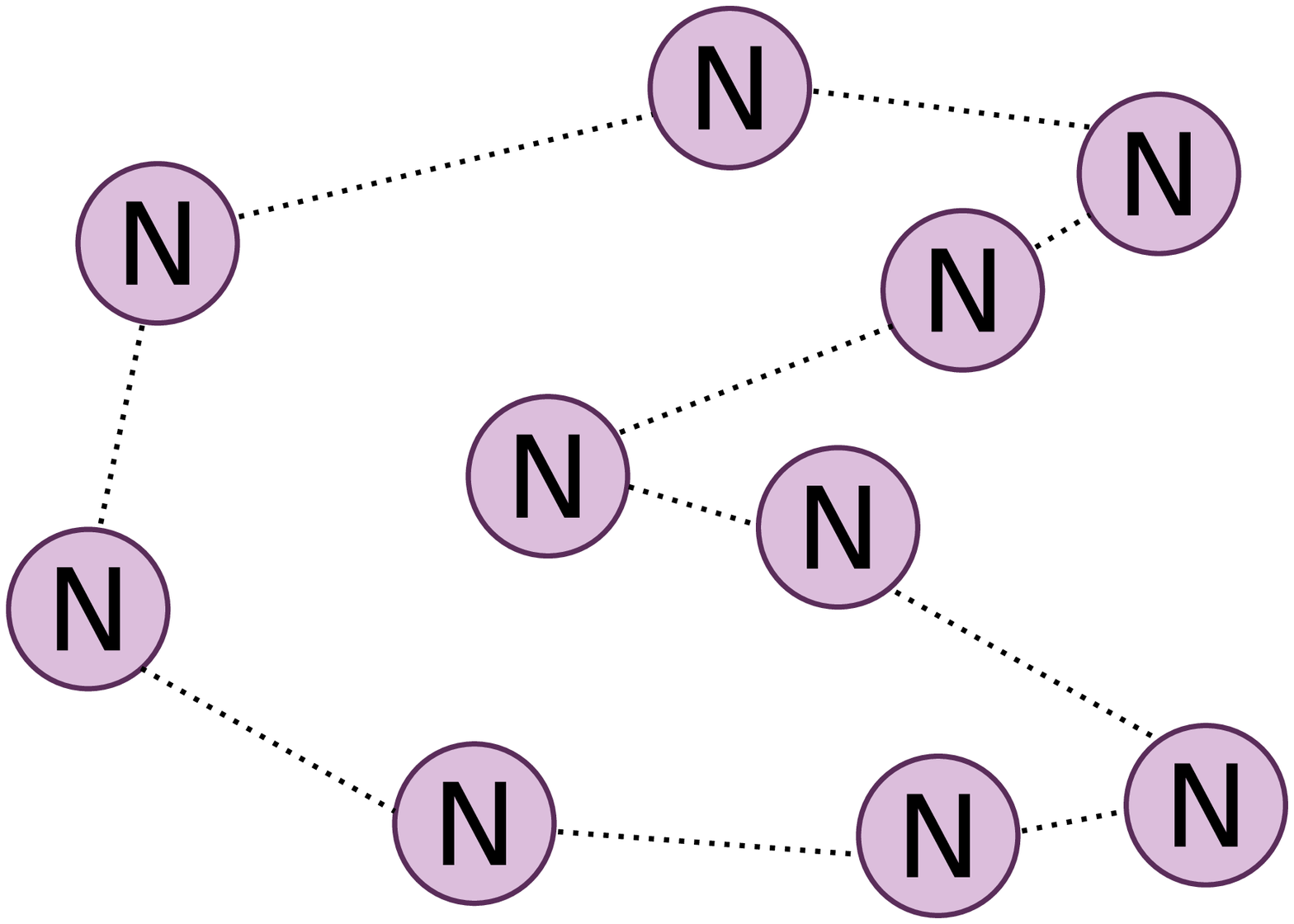}}}}%
\subfloat[]{
\framebox{\resizebox*{4cm}{!}{\includegraphics{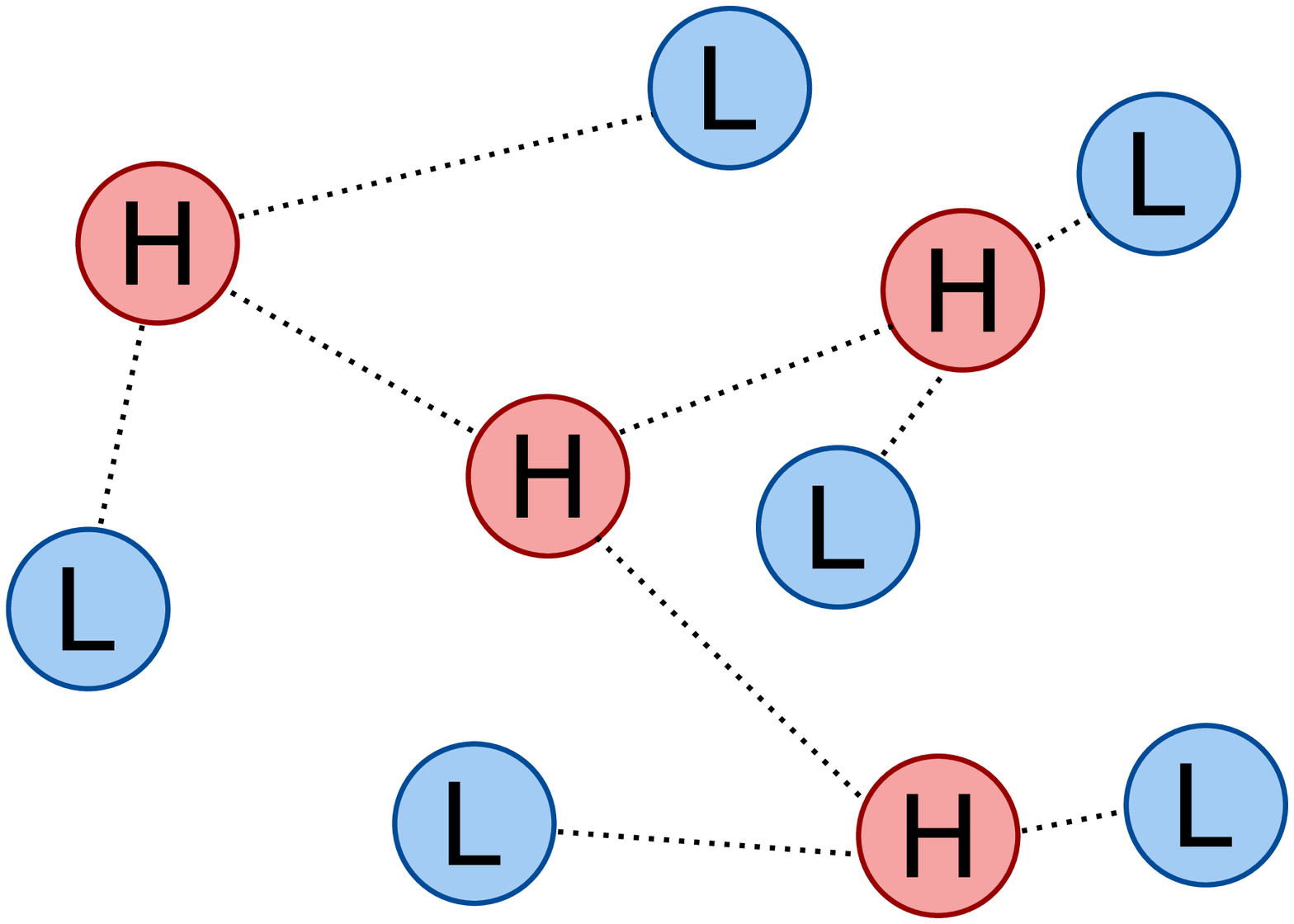}}}}%
\subfloat[]{
\framebox{\resizebox*{4cm}{!}{\includegraphics{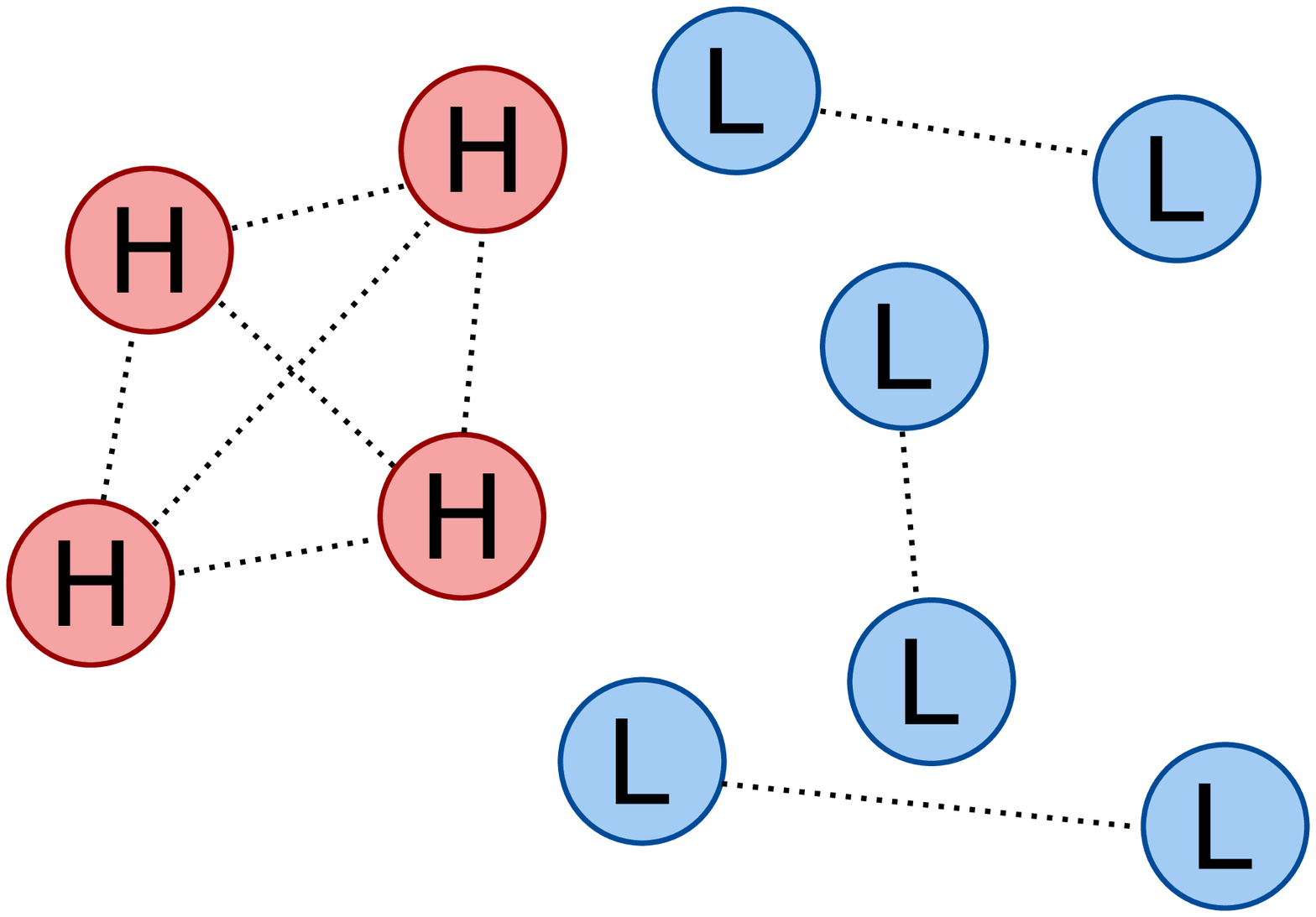}}}}%
\caption{(a) shows a regular graph in which every \underline{N}ode has degree
2. (b) shows a heterogeneous network where some nodes are
\underline{H}igh-degree with 3 links and others are \underline{L}ow-degree with
1 link, but there are connections between these two types of node. (c) shows a
highly degree-assortative heterogeneous network where high-degree and low-degree
nodes connect only to other nodes of the same type.}%
\label{fig:nets}
\end{minipage}
\end{center}
\end{figure}

We now move on to networks that contain more structure than ER random graphs.
Figure~\ref{fig:nets} shows three kinds of network: (a) a 2-regular graph in
which every node participates in two links; (b) a network with finite-variance
degree distribution and no preference for nodes to link to other nodes of a
similar degree; and (c) a network with finite-variance degree distribution and
a strong preference for nodes to link to other nodes of a similar degree.
We will consider each of these types of network in turn.

\subsection{Regular graphs}

The ER random graph model is something of a special case, since every link's
presence (or absence) is an independent chance event. One of the simplest
ways to introduce correlations between links is to constrain the node degree so
that every node has constant degree $n$. Such random graphs are called
$n$-\textit{regular}, and can be constructed in several ways; however it is
clear that the regularity condition means that each link's presence is not
independent of others'. A node that already has $n$ links cannot participate in
any more, while a node with 0 links must be allocated a further $n$. We
consider algorithms for dealing with this later, but for now let us think about
the giant component size for such graphs.

\subsubsection{Giant component size and percolation}

In order to consider this, we make an argument much like that made for ER
graphs above. Let $x$ be the probability that an individual in an $n$-regular
graph constructed in such a way as to avoid the presence of short loops is not
in the giant component. Then each of its $n$ neighbours must also not be in
the giant component. Writing $\tilde{x}$ for the probability that these neighbours
are themselves not in the giant component, we can write
\be
x = \tilde{x}^n \text{ ,} \qquad \tilde{x} = \tilde{x}^{n-1} \text{ .}
\label{reggc}
\ee
Then there are clearly three cases to consider.
\begin{enumerate}
\item[$n=1$:] So~\eqref{reggc} $\Rightarrow \tilde{x}=1=x$, and the network
is composed of isolated pairs of nodes.
\item[$n\ge 3$:] Then $x = \tilde{x} = 0$ satisfies~\eqref{reggc}, and all
nodes are in the giant component. 
\item[$n=2$:] This is the critical value for $n$. Going through a more careful
argument shows that long chains of nodes are formed, but the expected length of
the longest of these grows more slowly than the network size $N$.
\end{enumerate}
While it is possible to consider site percolation for these networks as a model
for vaccination, for simplicity let us consider how bond percolation affects
the result~\eqref{reggc} to give
\be
x = \left((1-p_b) + p_b \tilde{x}\right)^n \text{ ,} \qquad 
\tilde{x} = \left((1-p_b) + p_b \tilde{x}\right)^{n-1} \text{ .}
\label{regperc}
\ee
These equations do not have a simple analytic solution (although for small $n$,
the polynomials involved can be factorised) but are quick to solve numerically.

\subsubsection{Transmission dynamics}

\label{S:pw}

So what is the equivalent to the dynamical \SIR{} model~\eqref{sir} for
networks?  In general, the disease state of node $i$ is either $S$, $I$ or $R$.
We write $A_i$ to indicate this: $S_i=1$ if $i$ is susceptible; $S_i=0$
otherwise; and similar definitions hold for other states. Then we also use a
notation where:
\be
[A] = \sum_i A_i \text{ ,}\quad
[AB] = \sum_{i,j} A_i B_j G_{ij} \text{ ,}\quad
[ABC] = \sum_{i,j,k} A_i B_j C_k G_{ij} G_{jk} \text{ ,}
\ee
making use of the adjacency matrix $\mathbf{G}$ defined in~\eqref{Gdef} above.
We assume, as before, that infectious individuals recover at a constant rate
$\gamma$, but now infection does not happen homogeneously at rate $\beta$ as in
the simple \SIR{} model~\eqref{sir}.  Instead, susceptible-infectious pairs
$[SI]$ become infectious-infectious $[II]$ at rate $\tau$.  It follows that an
epidemic on a network obeys the exact, but unclosed, system of equations
\begin{align}
\dot{[S]} & = - \tau [SI] \text{ ,}&
\dot{[SS]} & = - 2 \tau [SSI] \text{ ,}& \nonumber \\
\dot{[I]} & = \tau [SI] - \gamma[I] \text{ ,}&
\dot{[SI]} & = \tau ( [SSI] - [ISI] - [SI] ) -\gamma [SI] \text{ ,}\nonumber \\
\dot{[R]} & = \gamma[I] \text{ ,}&
\dot{[SR]} & = -\tau [ISR] + \gamma [SI] \text{ ,}\nonumber \\
&&\dot{[II]} & = 2 \tau ( [ISI] + [SI]) -2 \gamma [II] \text{ ,}\nonumber \\ 
&&\dot{[IR]} & = \tau [ISR] + \gamma ( [II] - [IR]) \text{ ,} \nonumber \\
&&\dot{[RR]} & = \gamma [IR] \text{ .}
\label{pw}
\end{align}
One could, of course, keep writing down equations for the triples in terms of
higher-order structure, but it is better to make assumptions that allow us to
close these equations. For an $n$-regular graph, the typical choice is
\be
[ABC]\approx \frac{n-1}{n} \frac{[AB][BC]}{[B]} \text{ .}
\label{pwclose}
\ee
This follows from assuming that nodes of type $A$ and $C$ are multinomially
distributed about nodes of type $B$ with probabilities $[AB]/(n[B])$ and
$[CB]/(n[B])$ respectively~\cite{Dangerfield:2009}.  When first introduced,
this was really an assumption of Type III in the language of \S\ref{sec:intro},
above made just to get some purchase on the dynamical system.  It turns out,
however, that this assumption is numerically extremely accurate for
\SIR{} dynamics, and recent results suggest a formal proof for this
observation~\cite{Decreusefond}.

There are two relevant results that can be obtained from manipulation of the
closed dynamical system obtained by substitution of~\eqref{pwclose}
into~\eqref{pw}~\cite{Keeling:1999}. The first of these is the analogue
of~\eqref{linsir} above.  Early in the epidemic, 
\be
I(t) \propto e^{r t} \text{ , }\quad \text{where}\quad
r = (n-2)\tau - \gamma \text{ .}
\label{linpw}
\ee
Therefore, we recover~\eqref{linsir} if we hold $\beta = n\tau$ constant while
taking $n\rightarrow\infty$. It is also possible to manipulate the differential
equations in a similar (but much more algebraically complex) manner to that 
used to derive~\eqref{sreq}, which gives a result for the final proportion of
the population suscepible, $s$, as
\be
s = \left(1 - \frac{\tau}{\tau+\gamma} + \frac{\tau}{\tau+\gamma}
s^{(n-1)/n} \right)^n \text{ .} \label{pwfs}
\ee
This is clearly equivalent to~\eqref{regperc}, but with the probability of
transmission across a network link $\tau/(\tau+\gamma)$ taking the place of the
bond percolation probability $p_b$.  So while the algebra gets more complex,
for graphs that are correlated through having fixed degree, it is possible to
make a link between epidemic dynamics and network theory.

\subsubsection*{Biological insight}

When there is just one infectious individual on an $n$-regular graph, then the
rate of exponential growth of infectious individuals is $n\tau -\gamma$, so why
is there a factor of $-2$ in equation~\eqref{linpw}? The first thing to note is
that every non-initial infectious individual must have been infected by
someone, and that removes one from its potential pool of susceptibles,
explaining half of the $-2$. The other half must therefore arise because early
in the epidemic, the average infectious individual has also already infected
exactly one of its contacts. The practical implication of this is that an
epidemic on a regular graph will grow more slowly than its transmission and
recovery rates would suggest.

\subsection{The configuration model}

\begin{figure}
\begin{center}
\begin{minipage}{140mm}
\subfloat[]{
\resizebox*{7cm}{!}{\includegraphics{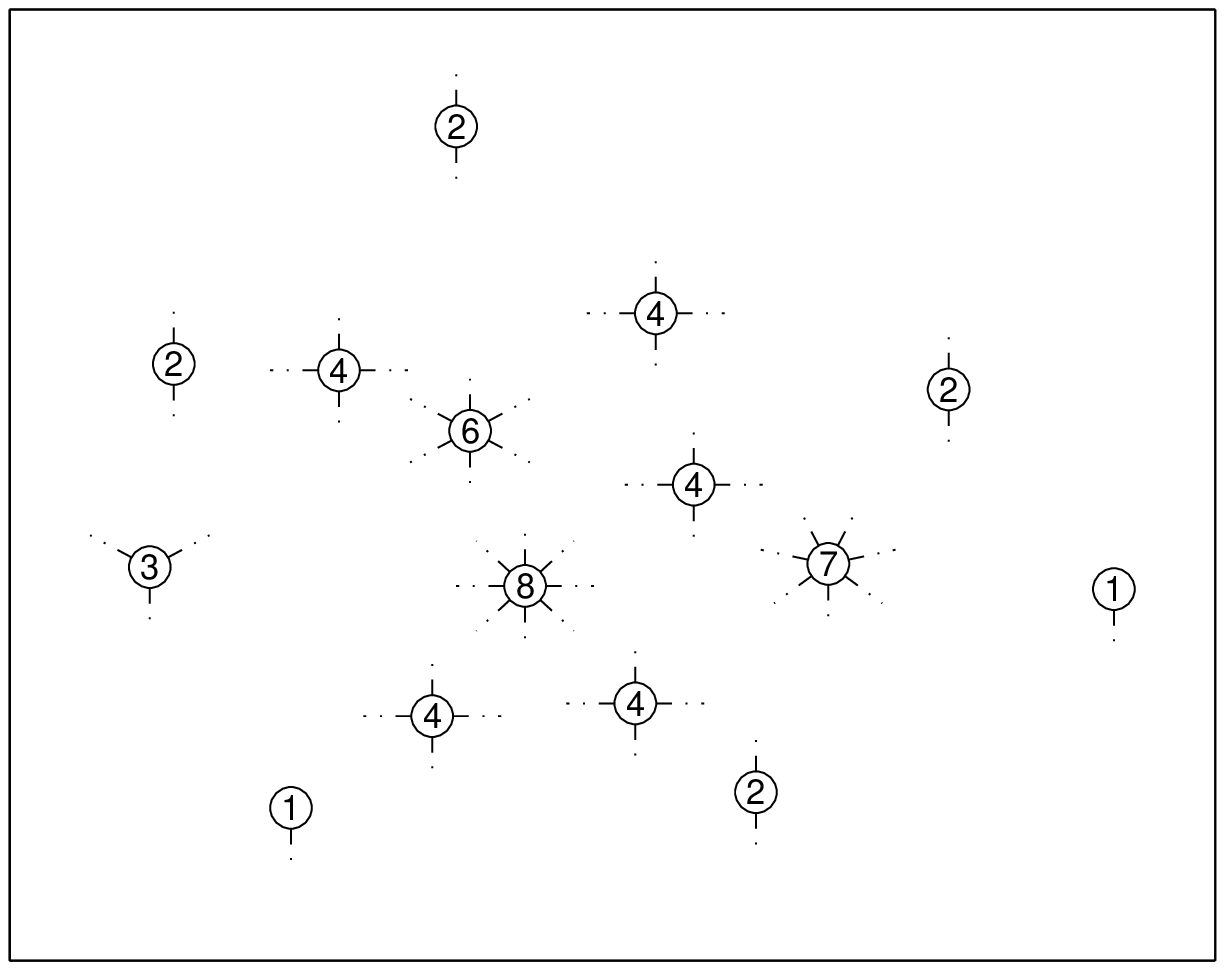}}}%
\subfloat[]{
\resizebox*{7cm}{!}{\includegraphics{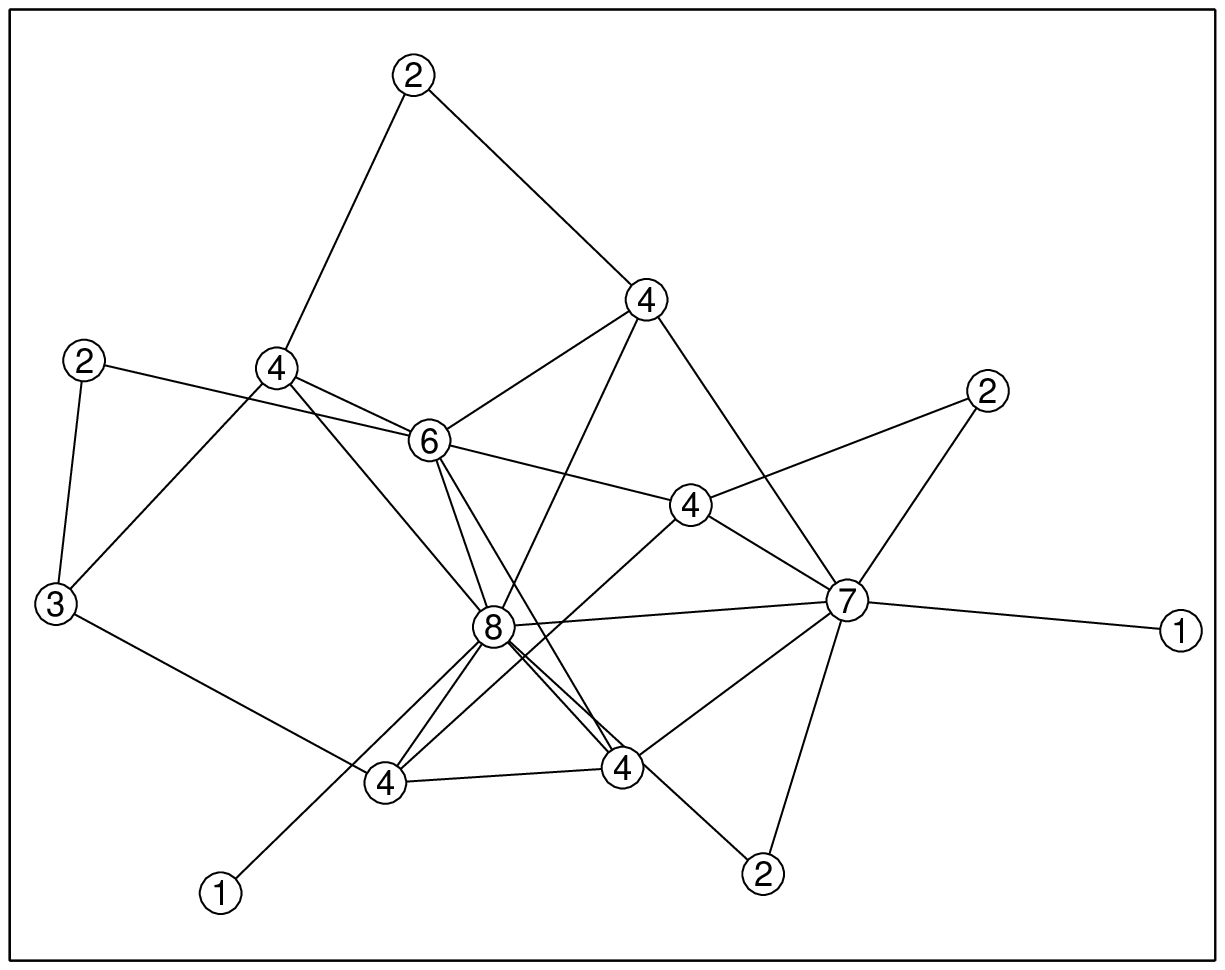}}}%
\caption{Construction of a CM network (a) individuals start with a degree,
and are given that number of stubs, then (b) stubs are paired at random
to form a network.}%
\label{fig:makecm}
\end{minipage}
\end{center}
\end{figure}

In the discussion above, I dodged the question of how to construct a regular
graph. The configuration model (CM)~\cite{Molloy:1995} provides a way to
construct networks with a given degree distribution; however first it is worth
considering what this might mean. Returning to Figure~\ref{fig:nets}, the
configuration model is designed to construct networks of types (a) and (b) --
these have few short, closed loops and no particular preference for connections
between nodes of similar degree. The method for doing this is shown in
Figure~\ref{fig:makecm}. Firstly, each node is given a number of `stubs', and
then these are paired up in a random order to give a network. This process can
lead to repeated links, and links that start and end on the same node, but for
most practical purposes these can be ignored.

\begin{figure}
\begin{center}
\begin{minipage}{120mm}
\centering
\subfloat[]{
{\resizebox*{6cm}{!}{\includegraphics{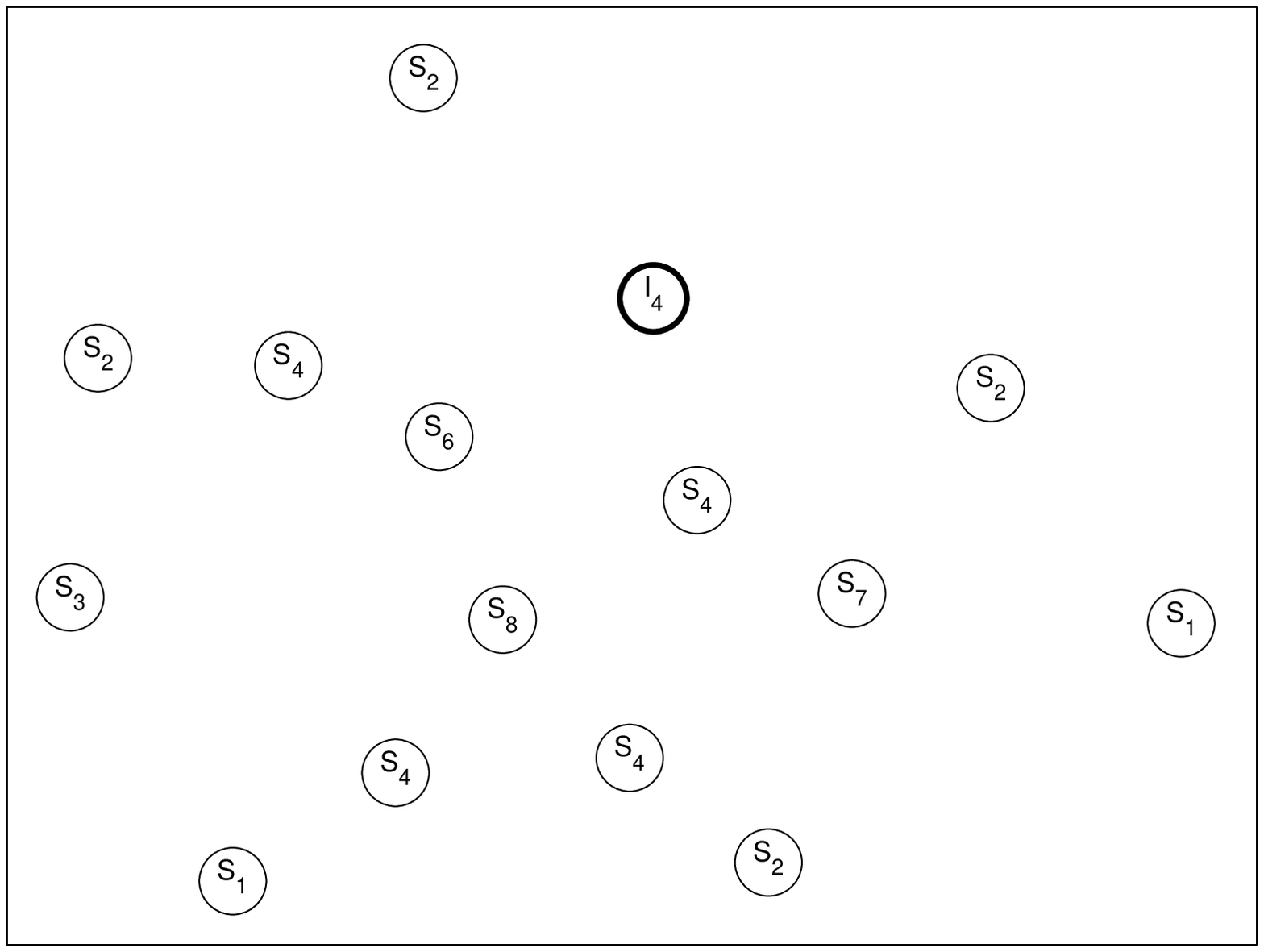}}}}%
\subfloat[]{
{\resizebox*{6cm}{!}{\includegraphics{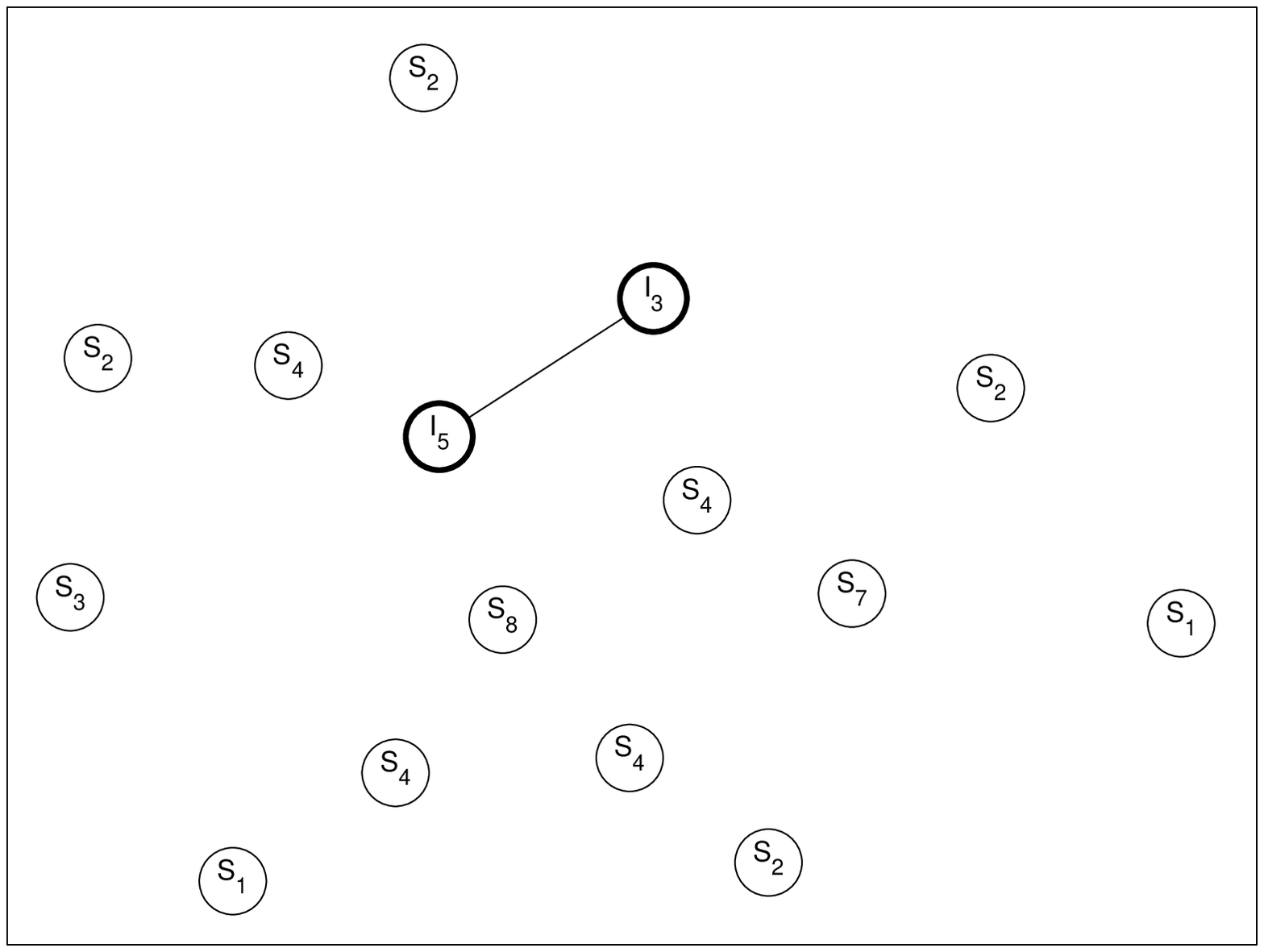}}}}\\%
\subfloat[]{
{\resizebox*{6cm}{!}{\includegraphics{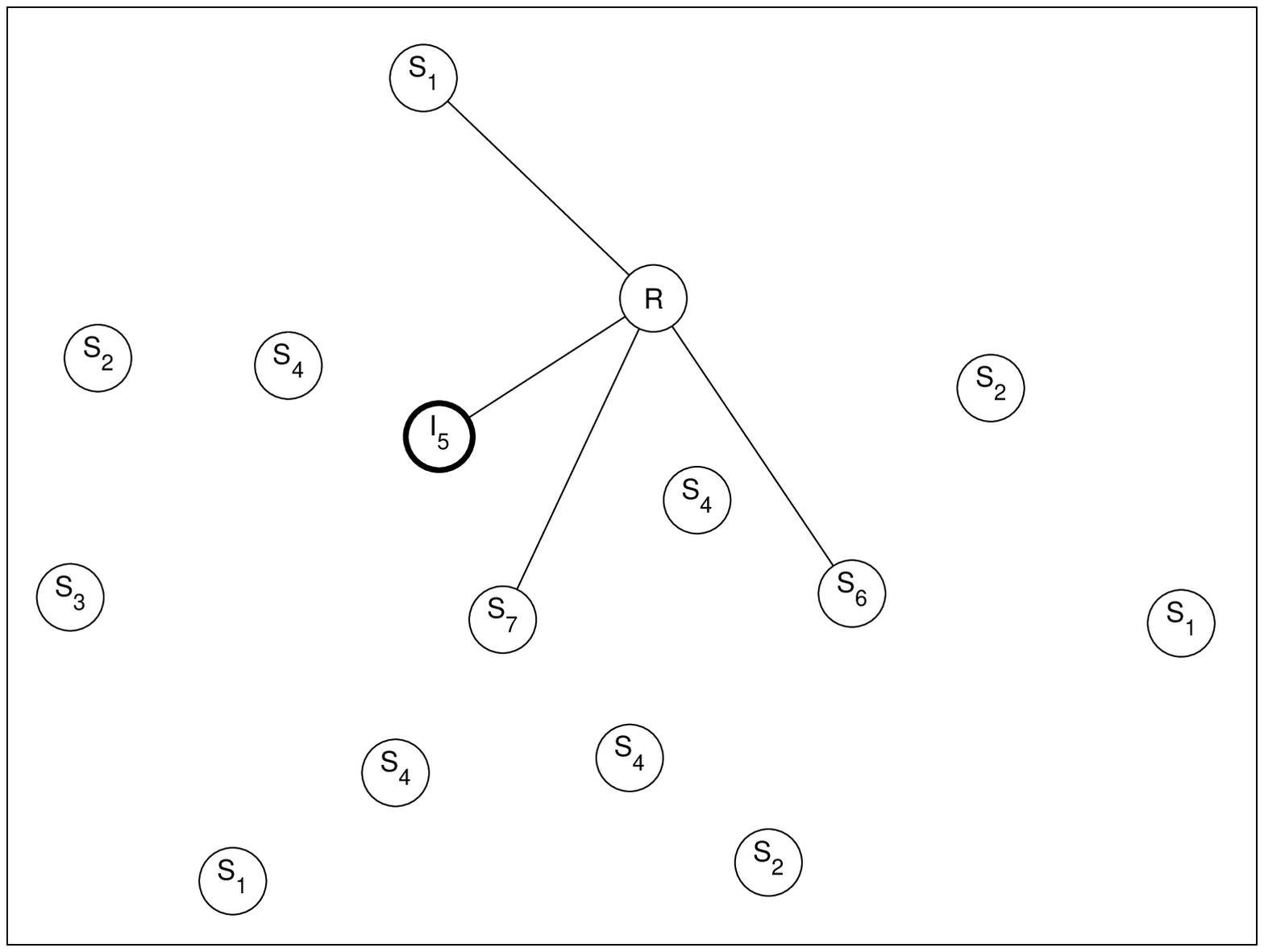}}}}%
\caption{The Ball \& Neal construction. Every individual is initially allocated
a number of stubs (a) An individual is the index case. (b) Infectious
individuals make links to other nodes proportionally to the target node's stub
number. Making a link to a susceptible individual makes that individual
infectious. Making a link reduces the stub number of both nodes in the link.
(c) Individuals recover at rate $\gamma$, at which point they use up their
remaining links at random. }%
\label{fig:makebn}
\end{minipage}
\end{center}
\end{figure}

\begin{figure}
\begin{center}
\begin{minipage}{60mm}
\resizebox*{6cm}{!}{\includegraphics{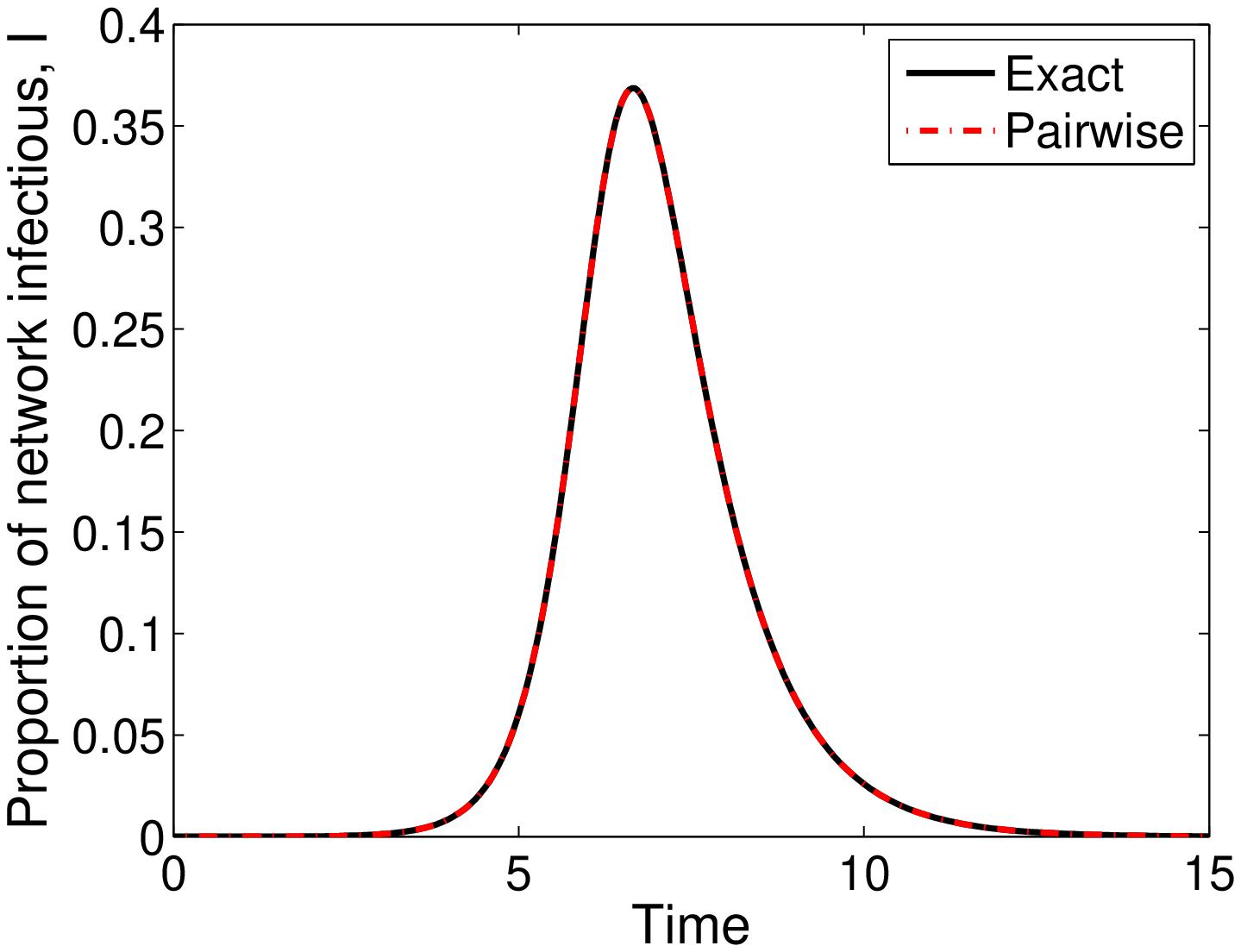}}%
\caption{Comparison of pairwise and exact models on a 3-regular graph.}%
\label{fig:pwex}
\end{minipage}
\end{center}
\end{figure}

In terms of epidemic dynamics on configuration model networks, a recent paper
by Ball \& Neal~\cite{Ball:2008} noted that these could be reconstructed if the
network is constructed at the same time as the epidemic. The construction
behind this is shown in Figure~\ref{fig:makebn}, and yields a closed system of
differential equations that we will call the BN model (a conceptually similar
but much lower dimensional set of equations was derived in~\cite{Volz:2008},
with recent results showing that this approach is also
exact~\cite{Decreusefond}). Simpler approaches can therefore be tested against
the BN model. Figure~\ref{fig:pwex} shows the typical results of doing this --
pairwise and related models are numerically indistinguishable from the BN
predictions.

There are therefore various ways to derive the following result, which was
prefigured in~\cite{Diekmann:2000}, and generalises~\eqref{linpw}:
\be
I(t) \propto e^{r t} \text{ , }\quad \text{where}\quad
r = \left(\text{mean}(k) + \frac{\text{var}(k)}{\text{mean}(k)}-2\right)\tau 
- \gamma \text{ ,}
\label{lincm}
\ee
where $k$ is the node degree.  This shows that an epidemic is only possible,
regardless of the rate of transmission compared to recovery, if
\be
\text{mean}(k) + \frac{\text{var}(k)}{\text{mean}(k)}>2
\text{ ,} \label{cmcrit}
\ee
which turns out to be the criterion for the existence of a giant component of
a CM network. We therefore retain the link between network topology (and hence
percolation) and epidemic dynamics.

\subsubsection*{Biological insight}

An interesting consequence of~\eqref{lincm} is that regardless of how small the
mean node degree is, a high variance can preserve a giant component and the
ability of a disease to spread.  This leads to highly connected individuals
playing a particularly important role in the spread of disease, and there is
some debate about whether an `80/20' rule holds for epidemics, with a small
fraction of the population causing the majority of transmission. Were this to
be the case, then interventions targeted at the highly connected individuals
alone could stop a disease spreading, although it is appropriate to be cautious
when proposing any new measure to control a disease.

\subsection{Assortativity}

The configuration model captures an important feature of real networks, namely
that some individuals have more connections than others. What it ignores is
the possibility that highly connected individuals may have a preference to
make links with other highly connected individuals. If we use notation like
the pairwise model, so that $[k]=N_k$ is the number of nodes of degree $k$,
and $[lm]$ is the number of links between a node of degree $l$ and a node of
degree $m$ in the network, then it is possible to quantify the preferences
of individuals through a symmetric correlation matrix $\mathcal{C}$:
\be
\mathcal{C}_{l,m} = \frac{[lm] \sum_k k[k]}{l[l]m[m]} \text{ .}
\ee
If every $\mathcal{C}_{l,m}=1$, then the correlations are consistent with the
configuration model. If $\mathcal{C}_{l,m}>1$ for similar values of $l$, $m$
and $\mathcal{C}_{l,m}<1$ for dissimilar values of $l$, $m$, then the network
is called \textit{assortative}; and a network is \textit{disassortative} if the
opposite relationship between $\mathcal{C}$ and the similarity of its indices
holds. Of course, there is much more information in a matrix than can be
encoded unambiguously in either a binary choice between assortativity and
disassortativity -- the ambiguity being essentially what one means by `similar'
-- but these remain useful concepts for thinking about epidemic networks.

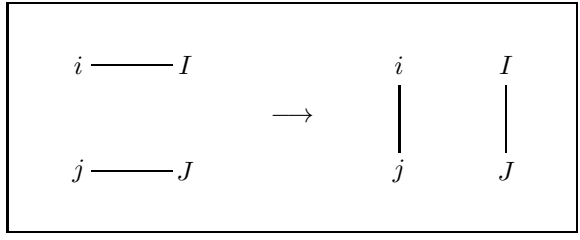
\begin{figure}
\begin{center}
\begin{minipage}{75mm}
\framebox{
\begin{picture}(200,80)(0,0)
\put (20,20){\makebox(0,0){$j$}}
\put (60,20){\makebox(0,0){$J$}}
\put (20,60){\makebox(0,0){$i$}}
\put (60,60){\makebox(0,0){$I$}}
\put (140,27){\line(0,1){25}}
\put (180,27){\line(0,1){25}}
\put (100,40){\makebox(0,0){$\longrightarrow$}}
\put (140,20){\makebox(0,0){$j$}}
\put (180,20){\makebox(0,0){$J$}}
\put (140,60){\makebox(0,0){$i$}}
\put (180,60){\makebox(0,0){$I$}}
\put (25,20){\line(1,0){30}}
\put (25,60){\line(1,0){30}}
\end{picture}
}
\caption{Rewiring / link-swapping move for construction of assortative
heterogeneous networks. This move is proposed by picking pairs of links
at random, and the proposed modification to the network is made or not
according to the standard Metropolis-Hastings rules.}%
\label{fig:rew}
\end{minipage}
\end{center}
\end{figure}

If we have a target $\mathcal{C}$ in mind, then Newman~\cite{Newman:2002}
suggested a method for producing a network with that correlation structure.
Starting with a CM network, link swaps are proposed as shown in
Figure~\ref{fig:rew}. Such a swap is then performed if
\be
\text{\texttt{rand}} < \frac{\mathcal{C}_{k_i,k_j}\mathcal{C}_{k_I,k_J}}%
{\mathcal{C}_{k_i,k_I}\mathcal{C}_{k_j,k_J}} \text{ ,}
\ee
where \texttt{rand} is a random number picked uniformly between 0 and 1. This
is a form of Metropolis-Hastings sampling, which should converge on a set of
networks with appropriate degree correlations given a large enough initial
network and sufficient computer time.

To model epidemics on such assortative networks, the paper~\cite{Eames:2002}
starts from a generalisation of~\eqref{pw} that indexes nodes by their degree:
\begin{align}
\nonumber
\dot{[S_k]} & = - \tau [S_kI] \text{ ,} &
\dot{[S_kS_l]} & = - \tau ( [S_kS_lI] + [S_lS_kI] )  \text{ ,} \\ \nonumber
\dot{[I_k]} & = \tau [S_kI] - \gamma[I_k] \text{ ,} &
\dot{[S_kI_l]} & = \tau ( [S_kS_lI] - [IS_kI_l] )
  - [S_kI_l] -\gamma [S_kI_l] \text{ ,} \\ \nonumber
\dot{[R_k]} & = \gamma[I_k] \text{ ,} &
\dot{[S_kR_l]} & = -\tau [IS_kR_l] + \gamma [S_kI_l] \text{ ,} \\ \nonumber
&&\dot{[I_kI_l]} & = \tau ( [IS_kI_l] + [IS_lI_k] )
 + [S_kI_l] + [S_lI_k] -2 \gamma [I_kI_l] \text{ ,} \\ \nonumber
&&\dot{[I_kR_l]} & = \tau [IS_kR_l] 
  + \gamma ( [I_kI_l] - [I_kR_l]) \text{ ,} \\ 
&&\dot{[R_kR_l]} & = \gamma ([I_kR_l] + [I_lR_k] ) \text{ ,} 
\end{align}
so $[A_k] = \sum_i A_i \delta_{k_i,k}$, where $\delta$ is the Kronecker delta,
is the number of nodes of degree $k$ in disease state $A$, and similarly for
pairs. Omission of a subscript index stands for an implicit sum, e.g.\
$[S_kI] = \sum_m [S_kI_m]$\;. As before, these equations are exact but
unclosed, and the moment closure proposed is
\be
[A_kB_lC_m] \approx \frac{(l-1)}{l} 
   \frac{[A_kB_l][B_lC_m]}{[B_l]} \text{ .}
 \label{fulltripleclose}
\ee
It is possible to manipulate the closed set of equations produced to derive
a further generalisation of~\eqref{linpw} and~\eqref{lincm}.
\be
I(t) \propto e^{r t} \text{ , }\quad \text{where}\quad
r = \left(\lambda(\mathbf{M}) - 1\right)\tau 
- \gamma \text{ ,}\quad 
(\mathbf{M})_{lm} = \frac{[lm](m-1)}{m[m]} \text{ ,}
\label{linas}
\ee
and $\lambda(\mathbf{M})$ is the dominant eigenvalue of matrix $\mathbf{M}$.
An equivalence between network theory and epidemic dynamics is maintained
here: $\lambda(\mathbf{M})>1$ is the condition for the existence of a giant
component.

\subsubsection*{Biological insight}

It is not immediately obvious what the dominant eigenvalue of a general matrix
looks like. But it turns out that in the same way that heterogeneous networks
can have low mean degree and still support an epidemic, it is possible for
assortativity to concentrate sustained transmission between highly connected
individuals even if the equivalent configuration model network would not
sustain an epidemic. An example of this is sexually transmitted diseases, where
much of the population could be in long-term partnerships, forming a set of
small components of size 2, while transmission is sustained amongst a `core
group'. In this scenario, the targeting of interventions at those with many
connections may well be suboptimal, since these may do little to reduce
prevalence in the core group while failing to halt transmission between the
core group and individuals on its periphery.

\section{Clustering}

\subsection{Small worlds}

To start with, let us define two network properties in terms of the adjacency
matrix $\mathbf{G}$ as defined in~\eqref{Gdef}. First, the clustering
coefficient $\phi$ is the number of triangles in the network divided by the
total number (closed and unclosed) triples:
\be
\phi = \frac{\sum_{i,j,k} G_{ij} G_{jk} G_{ki}}%
{\sum_{i,j,k} G_{ij} G_{jk} (1 - \delta_{ik})} \in [0,1] \text{ ,}
\ee
where $\delta_{ik}$ is the Kronecker delta.  Secondly, the shortest path length
between two distinct nodes $d_{ij}$ is the minimum number of links needed to
form an unbroken path between them:
\be
d_{ij} = \text{min} \{ p | (\mathbf{G}^p)_{ij} = 1 \} \text{ .}
\ee
The `small world' effect is essentially that observed networks of contacts
often have significant values of $\phi$, but low integer values of $d_{ij}$ --
your contacts are likely to contact each other, and you are probably at most six
handshakes (or even sneezes) away from the majority of people on Earth.

At first sight, this creates a paradox, because the easiest clustered networks
to visualise are lattices, which have large values of $d_{ij}$.  Watts and
Strogatz~\cite{Watts:1998} showed that this could be overcome, through the
introduction of a small number of random links to a lattice.  While this work
is widely (and correctly) perceived as solving an important conceptual problem,
other networks with the small worlds properties of significant $\phi$ and small
integer values for $d_{ij}$ have recently been proposed that are more realistic
for modelling epidemics.

\subsubsection*{Biological insight}

Historical epidemics like the Black Death spread over years through Europe at
walking pace, but in the modern age pandemics can cross continents in hours.
The low path lengths seen in `small world' networks explain this change.
People have kept the same household and local community contacts that they had
throughout history, but modern transportation means that business and leisure
travel can happen over previously unimaginable distances.  Short path lengths
have been a feature of all networks considered in this paper so far, but what
about the clustering in our local contacts? The precise impact of clustering on
epidemic dynamics is subtle, and cannot be condensed (yet) into a set of
straightforward biological insights. We shall consider instead some possible
routes to gain traction on the problem.

\subsection{Triangles and percolation}

The presence of an appreciable number of triangles in a network forms
a mathematical inconvenience for a rather subtle reason. To see why this
should be, let us consider two different scenarios and two different models.
Scenario I is an epidemic started from the middle node of an unclosed triple;
and Scenario II is an epidemic started from one node of a triangle. In Model
1, each infectious individual is equally transmissible and has a probability
$T<0.5$ of transferring infection to each of its contacts; and in Model 2,
half of infected individuals have zero probability of transmitting to each
contact and half have probability $\tilde{T}=2T$ of transmitting to each
contact.  Table~\ref{perctab} goes through all of the possibilities for
epidemics for these scenarios, and gives us the following results for
expected final epidemic size depending on scenario and model:
\begin{align}
R_\infty^{\mathrm{I},1} & = 1+2T \text{ ,}
& R_\infty^{\mathrm{I},2} & = 1+2T \text{ ,}
\nonumber \\
R_\infty^{\mathrm{II},1} & = 1+2T(1+T-T^2) \text{ ,}
& R_\infty^{\mathrm{II},2} & = 1+2T(1+T-2T^2) \text{ .}
\end{align}
Percolation therefore gives the correct epidemic final size for Model 1, or for
either model on an unclosed triple, but for Model 2 on a triangle we need a
process that correlates the presence of links with shared nodes meaning that
percolation is unsuitable.

\begin{table}
\begin{center}
{\begin{tabular}{l|l|l|l|l}
\textbf{Scenario} & \textbf{Epidemic} & $\mathbf{2^{\circ}}$ \textbf{cases} &
\textbf{Model 1 Prob.} & \textbf{Model 2 Prob.}\\
\hline
I: \mtIiv & \mtIi & 0 & $(1-T)^2$ & $\frac12(1+(1-\tilde{T})^2)$ \\
& \mtIii  & 1 & $T(1-T)$ & $\frac12\tilde{T}(1-\tilde{T})$ \\
& \mtIiii & 1 & $T(1-T)$ & $\frac12\tilde{T}(1-\tilde{T})$ \\
& \mtIiv & 2 & $T^2$ & $\frac12\tilde{T}^2$ \\
\hline
II: \mtII & \mtIIi & 0 & $(1-T)^2$ & $\frac12(1+(1-\tilde{T})^2)$ \\
& \mtIii & 1 & $T(1-T)^2$ & $\frac14\tilde{T}(1-\tilde{T})(2-\tilde{T})$ \\
& \mtIiii & 1 & $T(1-T)^2$ & $\frac14\tilde{T}(1-\tilde{T})(2-\tilde{T})$ \\
& \mtIIii  & 2 & $T^2(1-T)$ & $\frac14\tilde{T}^2(1-\tilde{T})$ \\
& \mtIIiii & 2 & $T^2(1-T)$ & $\frac14\tilde{T}^2(1-\tilde{T})$ \\
& \mtIIiv & 2 & $T^2$ & $\frac12\tilde{T}^2$ \\
\end{tabular}}
\end{center}
\caption{Table showing epidemic tree probabilities on I: an unclosed triple; and
II: a triangle; for Model 1: fixed infectious period; and Model 2: bimodal
infectious period. In the diagrams, a large circle corresponds to the initial
infectious individual, a solid line corresponds to a link that transmitted
infection, the absence of a line corresponds to no transmission of infection,
and a dotted line means a link whose role in transmission is not specified.}
\label{perctab}
\end{table}

\subsection{Local tree-like structure}

Given that percolation does not work in a straightforward manner on networks
with an appreciable clustering coefficient, a current research topic is to find
special kinds of clustered networks where there are short closed loops, but the
next level up in the network still looks `tree-like', meaning that if the
local, clustered structure is small enough to solve an exact epidemic model on
by going through a process like that in Table~\ref{perctab}, there is a chance
of piecing together solved local structures and solvable global connection
rules.

\begin{figure}
\begin{center}
\begin{minipage}{115mm}
\subfloat[]{
\framebox{\resizebox*{5cm}{!}{\includegraphics{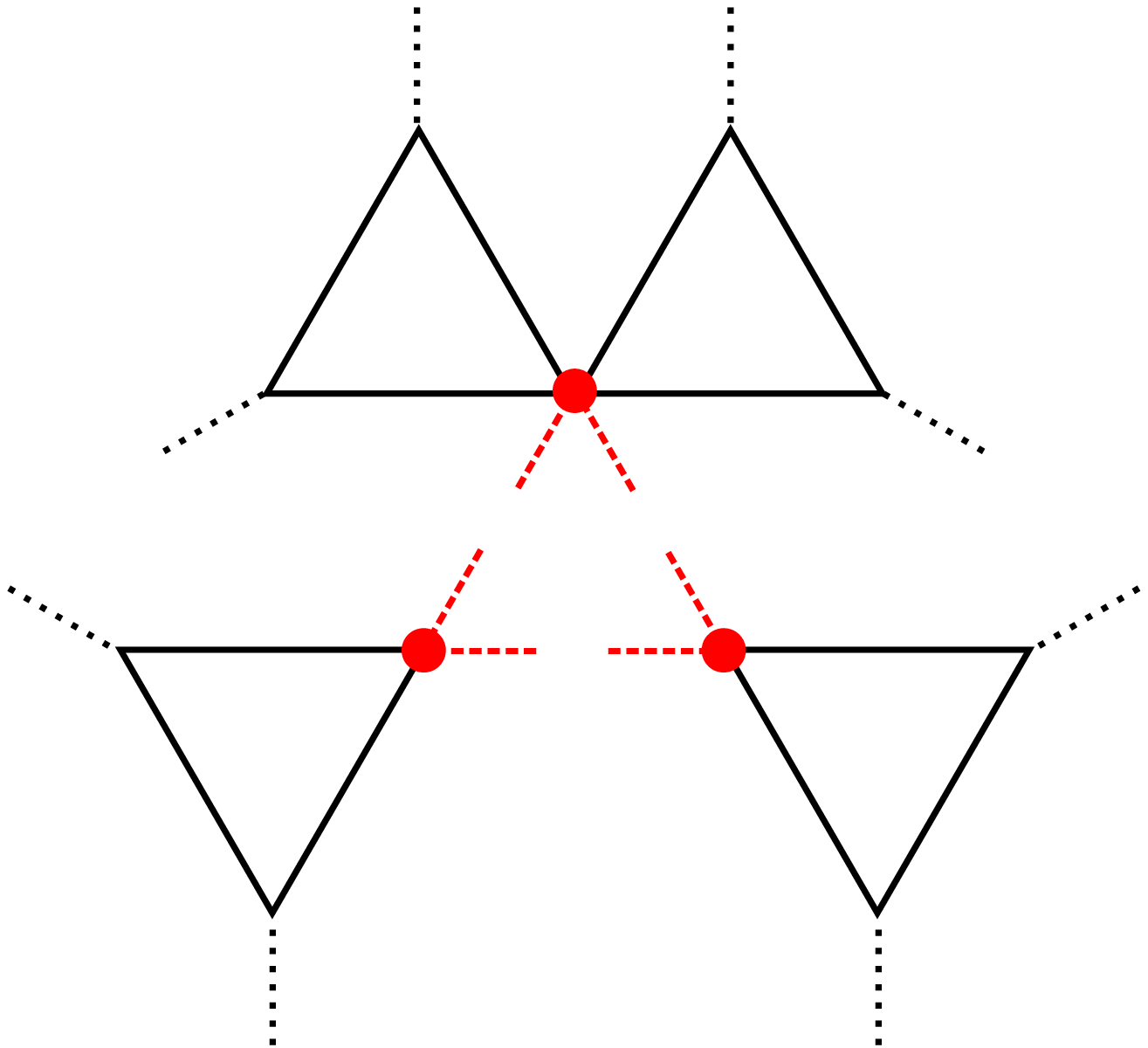}}}}%
\hspace{.5cm}
\subfloat[]{
\framebox{\resizebox*{5cm}{!}{\includegraphics{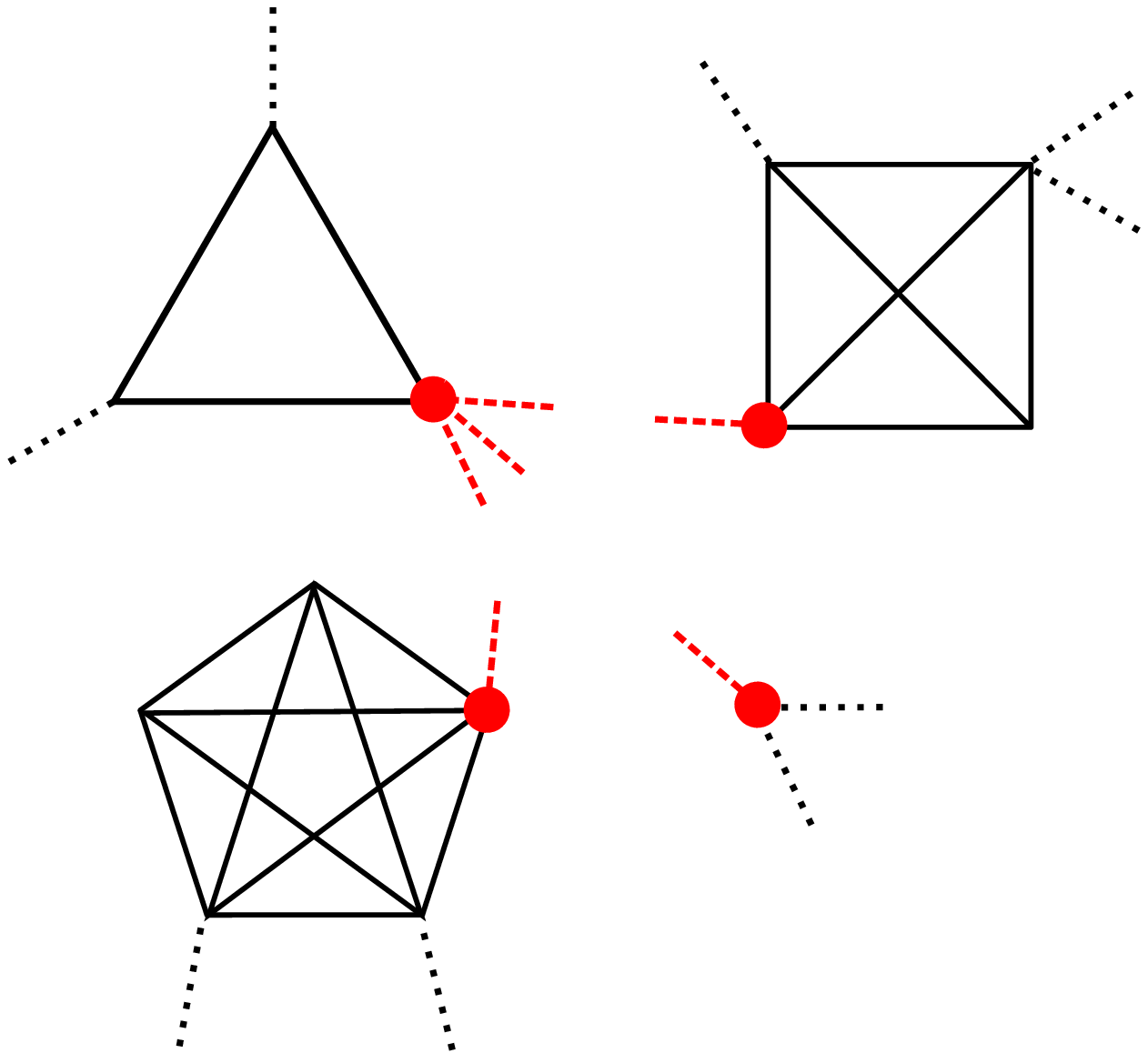}}}}%
\caption{(a) Construction step for a triangular configuration model network,
showing three nodes each with unmet demand coordinating to make a triangle. (b)
Construction step for a clique-based network, showing a node attached to a
3-clique with three stubs recruiting other stubs from a 4- and 5-clique with
one stub each and an isolated node with three stubs.}%
\label{fig:hg}
\end{minipage}
\end{center}
\end{figure}

Two such approaches are shown in Figure~\ref{fig:hg}.  The first of these,
shown in (a) is the triangle configuration
model~\cite{Newman:2009,Miller:2009}. In this generalisation of the CM, nodes
are assigned a demand for triangles in addition to normal links, and random
selections of three nodes at a time are made to satisfy this demand. The
second, shown in (b) is a clique-based model~\cite{Ball:2009,Ball:2010}, in
which nodes are placed in fully connected local subgraphs called
\textit{cliques}, then given an Configuration Model-like demand for global
links. Both of these constructions allow analytic results to be obtained, and
dynamics to be written down.

\subsection{Dynamical clustering}

The pairwise approach outlined in \S\ref{S:pw} above lends itself naturally to
consideration of epidemic dynamics on networks with appreciable clustering
coefficient. This is achieved by modification of the closure
relationship~\eqref{pwclose}.  The traditional clustered closure (sometimes
attributed to Kirkwood~\cite{Kirkwood:1942} and analysed for \SIR{} epidemics
in~\cite{Keeling:1999,House:2011}) is
\be
 [ABC] \approx \frac{n-1}{n} \frac{[AB][BC]}{[B]} \left( (1-\phi) 
  + \phi \frac{N}{n} \frac{[CA]}{[A][C]} \right) \text{ .}
\label{kwclose}
\ee
Several improvements to this closure have been suggested, including
some that are much more readily interpretable~\cite{House:2010,Rogers:2011},
but no network with significant $\phi$ has yet been found for which any given
closure is exact. Despite this, the system of equations is often good enough
for practical purposes~\cite{Ferguson:2001}. It is also possible to
use~\eqref{kwclose} together with~\eqref{pw} to derive analytic results; in
particular, an approximate linear correction to~\eqref{linpw}
is
\be
I(t) \propto e^{r t} \text{ , }\quad \text{where}\quad
r \approx
 {\tau} \left(
 (n-2)
 -\frac{2 (n-1) \left( 2(n-1)(n-2) \tau + n \gamma \right)}%
  {n^2 ((n-2) \tau +\gamma )}\phi  
 \right) - \gamma +O\left(\phi ^2\right) \text{ .} \label{linphir0}
\ee
In contrast to heterogeneity in degree distribution and assortativity above,
the inclusion of clustering reduces the potential of an epidemic to invade
a population at all values of transmission parameters.

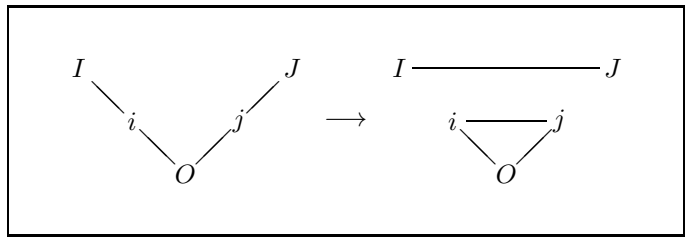
\begin{figure}
\begin{center}
\begin{minipage}{90mm}
\framebox{
\begin{picture}(240,80)(0,0)
\put (60,20){\makebox(0,0){$O$}}
\put (40,40){\makebox(0,0){$i$}}
\put (20,60){\makebox(0,0){$I$}}
\put (80,40){\makebox(0,0){$j$}}
\put (100,60){\makebox(0,0){$J$}}
\put (25,55){\line(1,-1){12}}
\put (43,37){\line(1,-1){13}}
\put (77,37){\line(-1,-1){13}}
\put (95,55){\line(-1,-1){12}}
\put (120,40){\makebox(0,0){$\longrightarrow$}}
\put (180,20){\makebox(0,0){$O$}}
\put (160,40){\makebox(0,0){$i$}}
\put (140,60){\makebox(0,0){$I$}}
\put (200,40){\makebox(0,0){$j$}}
\put (220,60){\makebox(0,0){$J$}}
\put (145,60){\line(1,0){70}}
\put (163,37){\line(1,-1){13}}
\put (197,37){\line(-1,-1){13}}
\put (165,40){\line(1,0){30}}
\end{picture}
}
\caption{`Big-V' rewiring / link-swapping move for construction of clustered
random graphs. This move is always accepted if it increases the clustering
coefficient.}%
\label{fig:bigv}
\end{minipage}
\end{center}
\end{figure}

While clustered pairwise models are attractive due to their relatively low
system dimension and number of parameters, the arbitrary nature of closure
proposals such as~\eqref{kwclose} is somewhat unsatisfactory and it would be
nice to have a better understanding of what makes a closure work (or not).  One
observation is that agreement is often best with $n$-regular graphs that have
had clustering introduced by the rewiring shown in
Figure~\ref{fig:bigv}~\cite{Bansal:2009,House:2010}, but understanding this
observation remains an active area of research.

\section{Concluding remarks}

This review has focused on methods for determining the conditions under which
an epidemic will take off in a population, paying particular attention to the
links between network theory and epidemiology. This helps to understand the
role that heterogeneity, preference and transitivity play in shaping the
transmission dynamics of human pathogens, and ultimately this understanding can
lead to the improvement of intervention strategies to control and mitigate the
burden of infectious disease.

My focus has been on the statistical physics technique of percolation, and the
dynamical systems technique of first-order differential equations.  A key
omission has been the contribution made by more mathematical researchers from
the field of probability theory, who have been able to derive many of the
results presented here in full rigour, and also shed light on the deep reasons
for the link between epidemics and networks, but this is done using techniques
that are not familiar to physicists and are so beyond the scope of this work.
Excellent monographs introducing applied probability approaches to random
graphs~\cite{Durrett:2007} and epidemics~\cite{Andersson:2000} are, however,
available.

For further reading, reviews that go into more detail on network epidemiology
include~\cite{Bansal:2007,Danon:2011}. I hope that this review, however, will
encourage readers with a background in physical sciences to invest the time in
reading more about this interesting field of biological research, where
techniques from physics can make an important contribution.

\section*{Acknowledgement} The author is funded by the UK Engineering and
Physical Sciences Research Council (EPSRC).

\end{document}